\documentclass[aps,twocolumn,floats,prl]{revtex4}
\usepackage{graphics,graphicx,epsfig}
\usepackage{amssymb}
\usepackage{epsf,epstopdf,wrapfig}

\begin{document}

\title{Information and fitness}

\author{Samuel F. Taylor,$^a$ Naftali Tishby$^b$ and William Bialek$^{a,c}$}

\affiliation{$^a$Joseph Henry Laboratories of Physics, $^a$Lewis--Sigler Institute for Integrative Genomics, and $^c$Princeton Center for Theoretical Physics, Princeton University, Princeton, NJ 08544\\
$^b$School of Computer Science and Engineering and Interdisciplinary Center for Neural Computation
Hebrew University, Jerusalem 91904, Israel}

 \begin{abstract}
 The growth rate of organisms depends both on external conditions and on internal states, such as the expression levels of various genes.  We show that to achieve a criterion mean growth rate over an ensemble of conditions, the internal variables must carry a minimum number of bits of information about those conditions. Evolutionary competition thus can select for cellular mechanisms that are more efficient in an abstract, information theoretic sense.  Estimates based on recent experiments suggest that the minimum information required for reasonable growth rates is close to the maximum information that can be conveyed through biologically realistic regulatory mechanisms.   These ideas are applicable most directly to unicellular organisms, but there are analogies to problems  in higher organisms, and we suggest new experiments for both cases.
 \end{abstract}

 \date{\today}

\maketitle 

Since Shannon's original work \cite{shannon_48,shannon_49} there has been the hope that information theory would provide not only a guide to the design of engineered communication systems but also a framework for understanding information processing in biological systems
\cite{attneave_54,barlow_61,more_neuro,ziv+al_06,tkacik+al_07}.
But there are (at least) two major obstacles in the way of  {\em any} effort to use information theoretic ideas in the analysis of biological systems.  First, Shannon's formulation of information theory has no place for the value or meaning of the information \cite{note1}, yet surely organisms find some bits more valuable than others.  Second, it is difficult to imagine that evolution can select for abstract quantities such as the number of bits that an organism extracts from its environment.  Both of these problems point away from general mathematical structures toward biological details such as the fitness or adaptive value of particular actions, the costs of particular errors, and the resources needed to carry out specific computations.  

The question of whether abstract information theoretic quantities can be connected to concrete costs and benefits is not new, nor is it specific to the biological context.  Fifty years ago, Kelly  asked
whether Shannon's  definition of information has a meaning outside the standard model of communication, and he showed that in simple models of gambling the rate at which one's winnings accumulate is bounded by the information (in bits) that one has about the outcome of the game \cite{kelly_56}.  Kelly's results generalize to the slightly more dignified setting of portfolio management \cite{cover+thomas_91}, and closely related ideas have emerged recently in thinking about phenotypic switching in bacteria \cite{bergstrom+lachmann_05,kussell+leibler_05}.  What these examples have in common is that the benefit  or growth (of investments, or of a bacterial population)  is a linear function of the control parameters (the fraction of the portfolio invested in each stock, or the fraction of organisms adopting a particular phenotype).  This linear framework is too restrictive, but Kelly's classical results encourage us to think that there may be some more general relationship between the information that an organism has about its environment and its growth rate or fitness.

\begin{figure}[bht]
\includegraphics[width = \linewidth]{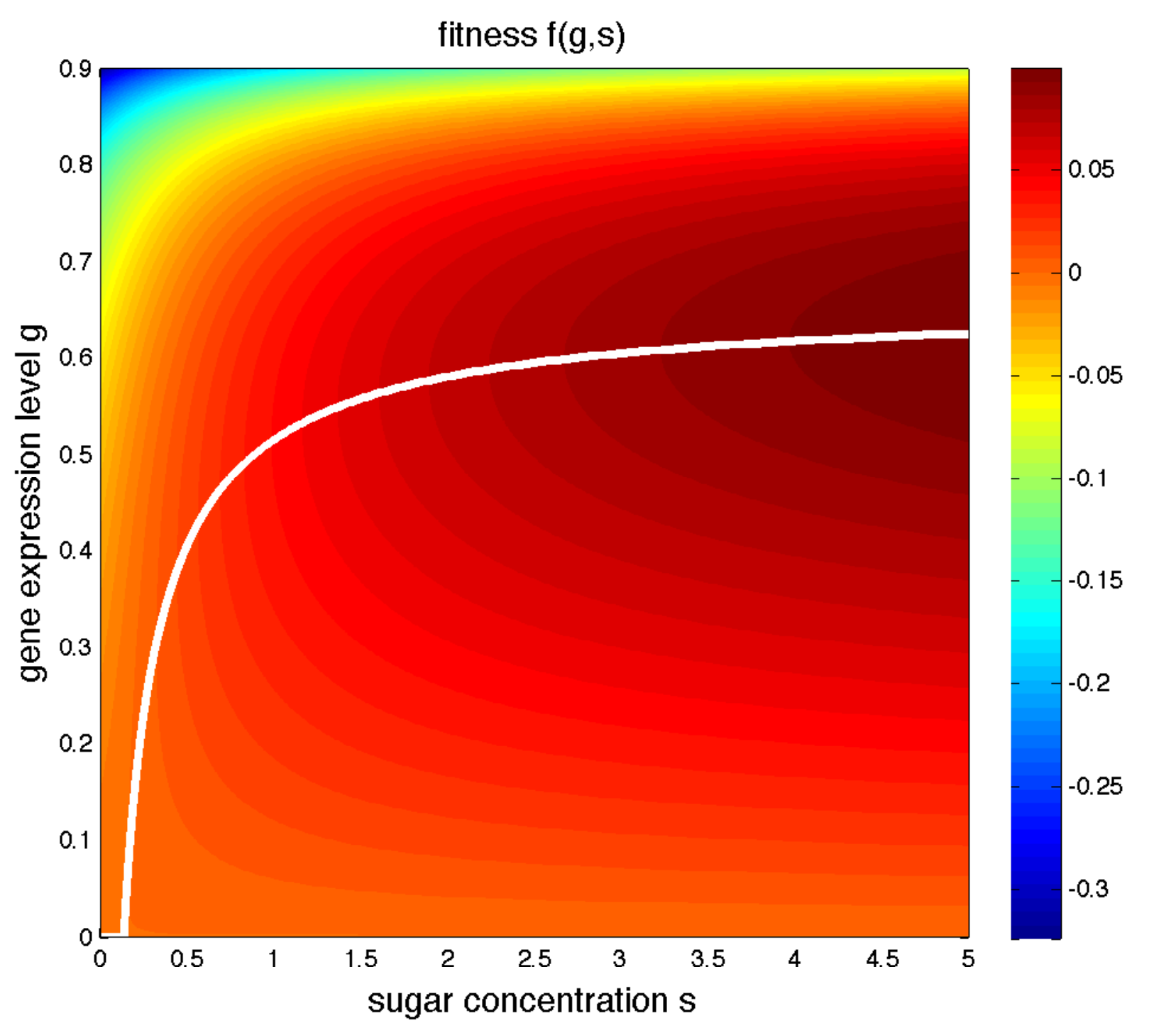}
\caption{Growth rate of {\em E coli} as a function of external sugar (lactose) concentration and the expression level of one gene ({\em lacZ}), as estimated in  Ref \cite{dekel+alon_05}.
 and summarized in their Eq (5). 
 Fitness is measured as a fractional difference from the growth rate when both the lactose concentration and {\em lacZ} expression levels are zero.  Sugar concentration is measured in units such that the half maximal benefit is reached at $s=1$, and expression level is measured in units of the maximum that the cell can maintain.  White line traces the optimal expression level for each value of $s$.}
\label{data}
\end{figure}

To be concrete, we consider single celled organisms in quasi--static environments, and discuss generalizations below. A bacterium lives in an environment described by a set of variables $\vec s \equiv s_1 , s_2 , \cdots , s_K$; in the simplest case, just one relevant environmental variable $s$ might specify the concentration of some limiting nutrient.  The fitness of the organism does not depend just on these environmental variables, but also on internal variables such as the expression levels of different enzymes involved in the metabolism of the available nutrients.  Let's refer to these variables as $\vec g \equiv g_1, g_2, \cdots , g_D$, and then the fitness of any particular organism in its environment is defined by some function $f(\vec g , \vec s )$ \cite{fit_note}.  This fitness function could be complicated---there are benefits to be gained from metabolizing particular nutrients, but achieving these benefits requires the appropriate expression levels of the relevant enzymes, and the expression of the proteins is itself a cost that lowers fitness.  Recent experiments attempt to map these different factors for the case of the {\em lac} operon in {\em E coli} \cite{dekel+alon_05}, resulting in an estimate of the fitness as a function of the environmental concentration of lactose ($s$) and the expression level of the lac proteins ($g$), shown in Fig \ref{data}.  The important point is perhaps not the detailed form found in particular experiments, but that we can imagine writing the fitness as depending on a combination of environmental and internal variables.

For any given set of environmental conditions there is some setting of the internal variables that provides for the maximum fitness.  If the organism could find this optimal operating point, then its internal state would be perfectly matched to the state of the environment.
Even if the system does not find this optimum, we can still think of the internal state as representing what the organism ``knows'' about the environmental variables.  To quantify this knowledge, we imagine that the organism will encounter, over its lifetime, a distribution $P({\vec s})$ of environmental conditions.  Given the state of the environment, organisms will adjust their internal state as best they can, but unless this process were (implausibly) noiseless, the result of the adjustment will be that the internal states are drawn from some probability dsitribution $P({\vec g} | {\vec s})$.  Thus if we were to take a snapshot, we would find individual cells with internal states $\vec g$ and their environments $\vec s$  drawn from the joint probability distribution $P({\vec g} , {\vec s}) = P({\vec g} | {\vec s}) P({\vec s})$.  Shannon then tells us that the internal state $\vec g$ provides information about the environment $\vec s$, and this information is
\begin{eqnarray}
I({\vec g}; {\vec s}) &=& \int d^K s \,d^D g  \, P({\vec g} , {\vec s})
\log_2
\left[  
{{P({\vec g} | {\vec s})}\over{P({\vec g})}} 
\right]\, {\rm bits,}\\
P({\vec g}) &=& \int d^K s \, P({\vec g} | {\vec s}) P({\vec s}) .
\label{Pg}
\end{eqnarray} 
The question is how this information content of the internal states relates to the fitness.

\begin{figure}[b]
\includegraphics[width =\linewidth]{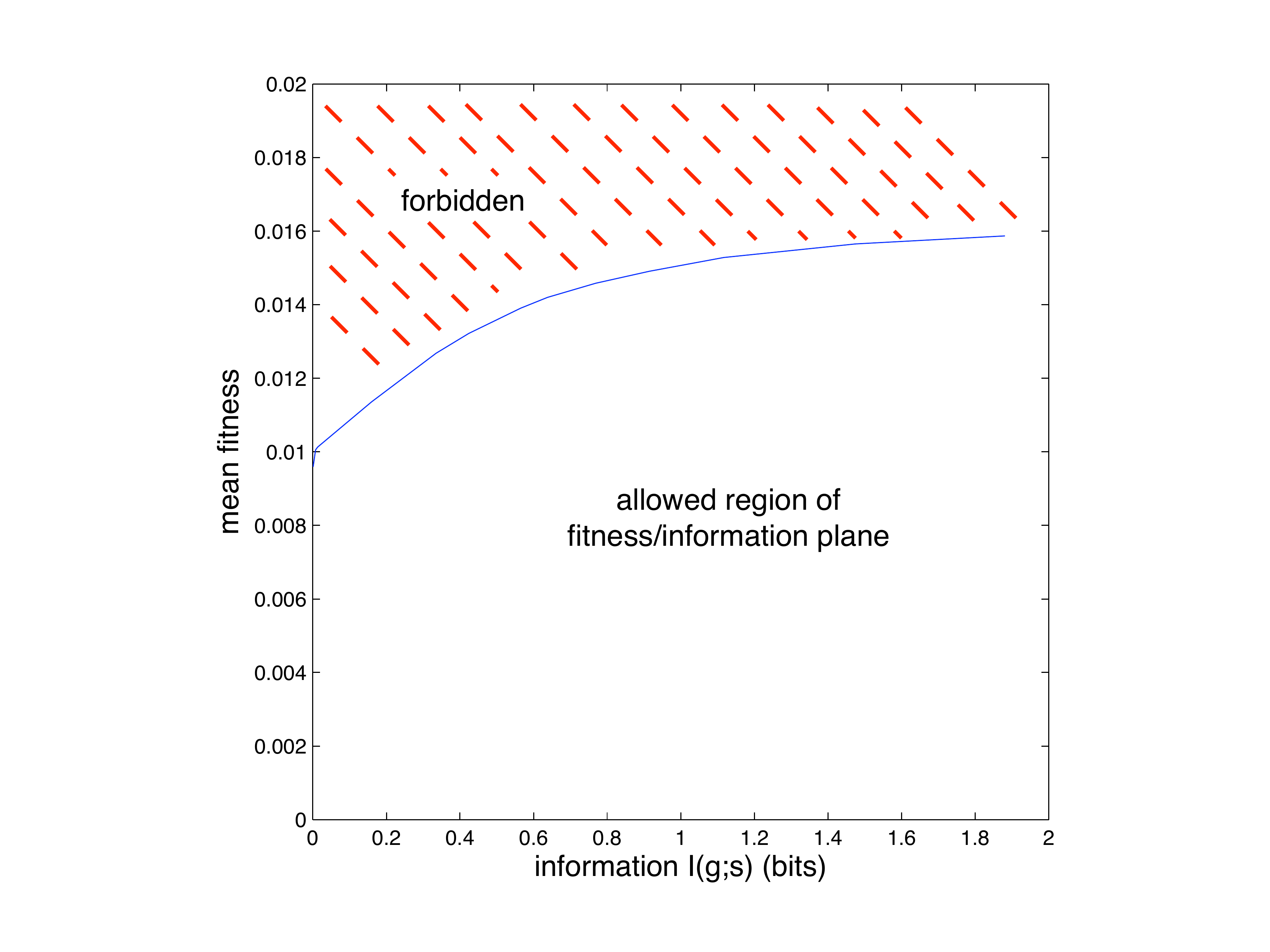}
\caption{Imagine mechanisms that tune the internal state of the organism in response to the state of the environment.  Each possible mechanism achieves a certain average fitness over the lifetime of the organism.  Depending on the precision of these mechanisms, the internal state will provide some amount of information  (in bits) about the state of environment.  Thus, each possible mechanism corresponds to a point in the plane relating the mean fitness $\langle f\rangle$ to the information $I({\vec g};{\vec s})$.  Not all points in this plane are physically possible:  there is a curve $I_{\rm min}(\langle f\rangle )$ that separates the allowed from the disallowed possibilities.  The example shown here is calculated for the fitness function in Fig \ref{data}, with a distribution of external states $P(s) \propto \exp(-2s)$.}
\label{schematic}
\end{figure}

Given the joint distribution of internal and external states, $P({\vec g}, {\vec s})$,  the average fitness over the organisms' experience in a distribution of environments is
\begin{eqnarray}
\langle f\rangle &=& \int d^K s \int d^D g \, P({\vec g} , {\vec s}) f(\vec g , \vec s )\\
&=& \int d^K s P({\vec s}) \int d^D g \, P({\vec g} | {\vec s}) f(\vec g , \vec s ) .
\end{eqnarray}
This is a linear function of the conditional distribution $P({\vec g}|{\vec s})$, while  the information $I({\vec g}; {\vec s})$ is a convex  function of the conditional distribution \cite{cover+thomas_91}.  Thus, if we consider all conditional distributions that lead to the same average fitness, then there is one which corresponds to the minimum amount of information; cf Fig \ref{schematic}.  The relationship between this minimal information and the mean fitness, $I_{\rm min}(\langle f\rangle )$, is analogous to the rate--distortion function in communication theory \cite{cover+thomas_91}.  We can also phrase this relation as $\bar f_{\rm max}(I)$, the maximum mean fitness that can be achieved given a certain amount of information.

The existence of the function $I_{\rm min}(\langle f \rangle)$ means that if organisms are to achieve a certain average level of fitness as they experience different environments, then the internal state of the organism $\vec g$ must provide a minimum amount of information about the relevant variables in the environment.  In this precise sense, achieving  a criterion level of fitness requires a minimum number of bits.  If evolution selects for greater fitness---as it does, almost by definition---then this selection continually raises the minimum number of bits that organisms need to represent about their environment.  Contrary to a widely held intuition, then, evolution {\em does} select for an abstract, information theoretic property.

The optimization problem in which we minimize the information $I({\vec g};{\vec s})$ at some fixed average fitness $\langle f \rangle$ has a simple formal solution,
\begin{equation}
P({\vec g} | {\vec s}) = {{P({\vec g}) }\over{Z({\vec s})}} e^{ \lambda f({\vec g} , {\vec s})},
\label{dist}
\end{equation}
where $\lambda$ is a Lagrange multiplier that   fixes the average fitness, and $Z({\vec s})$ serves to normalize each of the distributions $P({\vec g} | {\vec s})$.  The exponential of the fitness reminds us of the Boltzmann distribution, and one can think of $\lambda$ as being like an inverse temperature in statistical mechanics, biasing the distributions toward expression levels that insure higher fitness (lower energy) when $\lambda$ is larger (temperature is lower).  Equation (\ref{dist}) doesn't completely solve the problem because one must enforce consistency between $P({\vec g})$ on the right hand side and $P({\vec g} | {\vec s})$ on the left; that is, one must satisfy both Eq's (\ref{dist}) and (\ref{Pg}).  Fortunately these two equations can be combined into an iterative algorithm that converges \cite{BA}.

In Fig \ref{schematic} we show the results of a numerical compuation of the function $\bar f_{\rm max}(I)$, based on the fitness function in Fig \ref{data}. 
The precise results depend on the choice of the distribution of environmental conditions $P(s)$, but we have found that the scale and basic form of the function $\bar f_{\rm max}(I)$ are relatively robust so long as the distribution spans the range of sugar concentrations over which the optimal expressions levels actually vary.

The rate--distortion, or information--fitness function $I_{\rm min}(\langle f \rangle)$ is rather smooth and featureless.  Nonetheless, careful examination of the results illustrates several different points.  A significant fitness advantage ($\sim 1\%$, compared with the maximum possible $1.6\%$ across this ensemble of conditions) can be obtained by adjusting the gene expression level in ways that carry almost no information about the external world.  The distribution of expression levels under these conditions is still quite broad, corresponding to a population of organisms that uses (continuous) phenotypic diversity to survive under a range of conditions, but without a tight regulatory mechanism that links phenotype to the external world.

An even larger fraction of the available fitness advantage is accessible through mechanisms that use relatively little information, less than one bit. On the other hand, squeezing out the last $\sim 0.1\%$ of fitness advantage requires pushing well past one bit of information.  Put another way, organisms that could only implement a true switch--like control, in which expression is only ``on'' or ``off,'' would be at a small but measurable disadvantage in growth rate when averaged over a wide range of conditions. Organisms that have access to more than one bit of information thus could out--compete their one--bit cousins over thousands of generations.

We note that results based on the fitness function measured in Ref \cite{dekel+alon_05} are necessarily conservative.  Under the conditions studied in those experiments,  an infinite supply of the external sugar leads only to a $\sim 10\%$ advantage in growth rate over the case where the sugar is completely absent.  If we were to consider the case of a truly limiting nutrient, the overall scale of the fitness variations, and hence the curve $\bar f_{\rm max}(I)$, would thus be nearly ten times larger.  Thus, the difference between one bit and two bits would be $\sim 1\%$ in growth rate, which can be selected for on quite short time scales.

The scale of the information--fitness function also is interesting in comparison to what we know about the performance of real regulatory mechanisms \cite{tkacik+al_07b,caution}.  We recall that, because expression levels have a limited dynamic range and a finite amount of noise even under fixed conditions, the capacity of the expression level to convey information (about anything) is bounded.  With realistic parameters, based on recent experiments \cite{elowitz+al_02,ozbudak+al_02,blake+al_03,raser+oshea_04,rosenfeld+al_05,gregor+al_07b,tkacik+al_07c}, this capacity is less than three bits and more typically less than two bits \cite{tkacik+al_07b}.  Although we don't have enough data to reach a firm conclusion, these results certainly motivate the conjecture that the minimum information required to reach reasonable  levels of fitness is close to the maximum information that can be passed through real genetic regulatory elements.  

The precise form of the information--fitness function depends on 
the function $f({\vec g}, {\vec s})$, but the asymptotic behavior at high mean fitness is more nearly universal.  We can reach this limit by considering Eq (\ref{dist}) as the parameter 
$\lambda$ becomes large.  Then the distribution of expression levels becomes sharply peaked around the optimum ${\vec g}_{\rm opt}({\vec s})$ for each set of external conditions; the form of this distribution becomes approximately Gaussian with a width inversely proportional to $\lambda$.  Taking this Gaussian approximation seriously, it is straightforward to compute the information  and the mean fitness; we find
\begin{equation}
I_{\rm min}(\langle f\rangle )  = I_0 + {{D}\over 2}\log_2 \left[
{{D \langle f\rangle_{\rm max}}\over{2(\langle f \rangle_{\rm max} - \langle f \rangle)}}\right] + \cdots ,
\label{asymp}
\end{equation}
where $I_0$ is a constant independent of the mean fitness, $\langle f \rangle_{\rm max}$ is the maximum mean fitness obtainable by an organism that has perfect information about its environment, and $\cdots$ denotes terms which become relevant at lower fitness.  The details of the function $f({\vec g}, {\vec s})$ are buried in the constant $I_0$, but the way in which the minimum information grows as the organism approaches its maximal mean fitness
depends on the number of genes $D$ the cell has to control, independent of details.

We have assumed, for simplicity, that variations are slow, so we can write the fitness as a function of internal and external states at the same instant of time.  A more realistic analysis would take account of the fact that current values of internal control variables  interact with external conditions in the future,  so that the information which controls the achievable level of fitness is predictive information \cite{bialek+al_01,seattle}.  We can also generalize to consider behaviors in complex multi--cellular organisms; the analog of the information--fitness relation then states that behaviors which collect some criterion level of reward across an ensemble of conditions must be guided by neural representations which carry a minimum amount of information about these conditions.

It is possible to measure, in real time, both growth rates and expression levels of particular genes in individual unicellular organisms \cite{exps}.  Repeating such experiments under varying external conditions should allow estimates of the fitness function $f({\vec g},{\vec s})$ with single cell resolution, the mean growth rate under given conditions, and the mutual information between internal and external variables. Thus we could locate the organism's performance  in the information--fitness plane of Fig \ref{schematic}, and also see how close it comes to the limiting curve ${\bar f}_{\rm max}(I)$.  For neural systems, if we have a motor control task in which there is a good model of the underlying mechanics 
 \cite{motor}, analogous experiments  would compare the information available in central neural representations with the minimum required to achieve observed levels of reward under variable conditions. 

To summarize, achieving a criterion level of fitness or reward across a distribution of  conditions always requires an internal representation of the world that captures some minimum number of bits.  Qualitatively, this means that evolutionary competition will drive an increase in this information capacity.  Quantitatively, in the case of unicellular organisms, the minimum information required for reasonable levels of fitness is close to the maximal information that can be transmitted through known genetic regulatory mechanisms.  Finally, this general picture suggests experiments which could map the information/fitness tradeoff in a wider variety of systems, and locate the performance of real organisms in relation to the information theoretic optimum.

\acknowledgments{This work was supported in part by  NSF Grants IIS--0423039 and PHY--0650617,  by NIH Grant P50 GM071508, and by the Swartz Foundation.}


\begin{thebibliography}{99}
\bibitem{fit_note} 
Strictly speaking we should distinguish between the growth rate of individuals and their fitness relative to other organisms in the same environment, but we'll use ``growth rate'' and ``fitness'' as synonyms for simplicity.
\bibitem{caution}
Information could be also acquired passively, since organisms with the `correct' internal states will grow faster.  It seems difficult to make general statements about this effect, which depends on the dynamics with which different states are explored.
\end{thebibliography}
\end{document}